# Revealing Fundamentals of Charge Extraction in Photovoltaic Devices Through Potentiostatic Photoluminescence Imaging

Lukas Wagner[1,2,*], Patrick Schygulla[1,2], Jan Philipp Herterich[1,3], Mohamed Elshamy[1], Dmitry Bogachuk[1,2], Salma Zouhair[1,4], Simone Mastroianni[1,3], Uli Würfel[1,3], Yuhang Liu[5,*], Shaik M. Zakeeruddin[5], Michael Grätzel[5], Andreas Hinsch[1,*], Stefan W. Glunz[1,2]

[1] Fraunhofer Institute for Solar Energy Systems ISE, Heidenhofstraße 2, 79110 Freiburg, Germany.
Email: lukas.wagner@ise.fraunhofer.de, andreas.hinsch@ise.fraunhofer.de

[2] Laboratory for Photovoltaic Energy Conversion, University of Freiburg, Emmy-Noether-Straße 2, 79110 Freiburg, Germany.

[3] Freiburg Materials Research Center FMF, University of Freiburg, Stefan-Meier-Straße 21, 79104 Freiburg, Germany.

[4] Abdelmalek Essaadi University, FSTT, Thin Films & Nanomaterials Lab, 90000 Tangier, Morocco.

[5] Laboratory of Photonics and Interfaces (LPI), Department of Chemistry and Chemical Engineering, École Polytechnique Fédérale de Lausanne, Lausanne CH-1015, Switzerland.
E-mail: yuhang.liu@epfl.ch

## Abstract
The photocurrent density-voltage ($J(V)$) curve is the fundamental characteristic to assess opto-electronic devices, in particular solar cells. However, it only yields information on the performance integrated over the entire active device area. Here, a method to determine a spatially resolved photocurrent image by voltage-dependent photoluminescence microscopy is derived from basic principles. The opportunities and limitations of the approach are studied by the investigation of III-V and perovskite solar cells. This approach allows the real-time assessment of the microscopically resolved local $J(V)$ curve, the steady-state $J_{sc}$, as well as transient effects. In addition, the measurement contains information on local charge extraction and interfacial recombination. This facilitates the identification of regions of non-ideal charge extraction in the solar cells and enables to link these to the processing conditions. The proposed technique highlights that, combined with potentiostatic measurements, luminescence microscopy turns out to be a powerful tool for the assessment of performance losses and the improvement of solar cells.

## Introduction

The current density - voltage ($J(V)$) curve builds the basis of any further characterization and improvement of solar cells and other opto-electronic devices. This characteristic provides a broad range of information on the fundamental mechanisms such as charge carrier generation, recombination, and transport losses. Conventionally, the $J(V)$-curve is measured by illuminating the full device and externally probing the current flowing between the electrodes with a source meter. This approach therefore only provides information integrated over the entire device area. A range of techniques have been established for spatially resolved analysis of photovoltaic devices[1,2] such as light beam induced current mapping (LBIC),[3–5] lock-in thermography (LIT),[6–8] photoluminescence (PL) imaging,[3,9–23] and electroluminescence (EL) imaging.[24] However, as further discussed in the Supporting Information (SI), these techniques can either probe only a portion of the local $J(V)$ characteristic or require numerical fitting to a postulated diode model. Especially for emerging PV technologies like perovskite solar cells (PSC) it is crucial to gain spatially resolved fundamental information about the charge transport and the extraction phenomena on microscopic as well as on macroscopic scale to further enhance the power conversion efficiency and long-term stability of these cells. Therefore, novel advanced characterization techniques are necessary.

The physics of optoelectronic devices can be divided into fundamental operation processes. For photovoltaic power generation, these are the photo-generation of charge carriers by the absorption of photons, the transport to and separation of the photogenerated electrons and holes at charge extraction layers (CELs), and the extraction of the charges into the outer device terminals (cables). The three processes have been described by so-called reciprocity relations by Donolato,[25] Rau,[26] and Wong and Green,[27] respectively. To each of these processes, fundamental loss mechanisms can be assigned, namely optical and absorption losses, radiative and non-radiative recombination mechanisms, and transport losses.

In this work, we introduce a method based on electrical bias dependent photoluminescence imaging that allows the immediate determination of the charge extraction efficiency of photo-generated charge carriers. The approach makes use of the observation that the difference of two photoluminescence images at different bias voltages yields direct spatially resolved information on the local $J(V)$. We propose to refer to this method as local charge extraction analysis by "Potentiostatic Photoluminescence Imaging" (PPI). We outline the theoretical principles of the PPI approach and experimentally demonstrate the validity by means of a close-to-ideal GaInAsP solar cell. The potential to identify morphological features of benign and poor charge extractions are investigated by means of high-efficiency (>20 %) perovskite solar cell investigating both microscopic images of the local short-circuit current $J_{sc}$ in steady state as well as local $J(V)$ curves. Based on these results, we demonstrate how distinct charge extraction signatures of the PPI method can be linked to physical properties of the perovskite layer, the charge extraction layers, and the respective interfaces.

## Theory

The PPI method is based on a detailed balance approach, considering that in steady state, the current density measured at the outer solar cell contacts, $J(V)$, can be expressed by the current density of the photo-generated charge carriers, $J_{gen}$, and the (internal) recombination in the perovskite layer and at respective interfaces, $J_{rec}(V)$,

$$J(V) = J_{gen} - J_{rec}(V). \qquad (1)$$

Here, we assume that in the considered operation range (0 V ≤ V ≤ $V_{oc}$), the internal generation current $J_{gen}$ is not affected by the applied bias. Moreover, without loss of generality, we assume that the photon flux to illuminate the sample is constant over time and spatially homogeneous.

The internal recombination processes comprise non-radiative and radiative components, $J_{n.r.}$ and $J_{rad}$, respectively: $J_{rec} = J_{n.r.} + J_{rad}$. They can be related to each other by

$$J_{n.r.}(V) = k(V) \cdot J_{rad}(V). \qquad (2)$$

Equation (2) is generally valid if $k$ is a function of the applied bias, $k = k(V)$. Here, we show that, if the probed sample has a diode ideality factor close to one and displays negligible resistive losses (high shunt resistance, low series resistance), then it is justified to assume a linear relationship between $J_{n.r.}(V)$ and $J_{rad}(V)$ and hence $k$ = const., which we will assume in the following. Empirical indications for this circumstance in perovskite solar cells have already been presented by Stolterfoht and coworkers.[23] A detailed theoretical discussion of this assumption as well as a general expression are presented in the Supporting Information (SI section B).

Now, Equation (1) can be expressed as

$$J(V) = J_{gen} - (1 + k) \cdot J_{rad}(V). \qquad (3)$$

$J_{rad}(V)$ can be related to the signal of a photodetector $S_{PL}(V)$ by

$$S_{PL}(V) = c \frac{J_{rad}(V)}{e}, \qquad (4)$$

where $c$ describes the probability that photons created by radiative recombination enter the detector area and are translated into a detector signal.

Using Equation (3), and (4), we can now relate the electrical photocurrent to the difference between the voltage-dependent PL intensity PL($V$) and the PL at open circuit. By normalizing the term, the expression becomes independent of setup-specific factors which makes the technique independent from elaborate calibration measures. We find that

$$\frac{S_{PL}(V_{oc}) - S_{PL}(V)}{S_{PL}(V_{oc})} = \frac{J(V)}{J_{gen}}. \qquad (5)$$

This relation shows that photoluminescence microscopy can be used to derive spatially as well as time-resolved images of the local $J(V)$ performance of a photovoltaic device. Two applications are especially interesting: First, by recording only two PL images, one at open circuit and one at short circuit, the image of the local short-circuit photocurrent density $J_{sc}$ can be derived. Secondly, by recording PL images at various voltages, the local $J(V)$ of specific spots on the cell can be investigated. These approaches are investigated in the following. Thereby, the $J(V)/J_{gen}$ results determined by PL microscopy will be denoted as $J(V)/J_{gen}|_{PL}$.

## Results and discussion

### Experimental validation of the PPI approach

**III-V-based photovoltaic device**

Many III-V compound semiconductors, such as gallium-arsenide (GaAs), indium-phosphide (InP), or combinations like $Ga_xIn_{1-x}As_yP_{1-y}$, have a direct band gap. Hence, for III-V semiconductor crystals with a low defect density radiative recombination is the dominant recombination mechanism. In addition, the high absorption coefficient allows using thin absorber layers in the order of a few micrometers. This fact along with sufficiently high carrier mobilities and lifetimes promises high power conversion potential of this material class in photovoltaic applications.[28,29] III-V compound semiconductors can be grown epitaxially with high crystal quality using for instance metalorganic vapor phase epitaxy (MOVPE). The highest efficiencies of both single- and multi-junction solar cells under the AM1.5g solar spectrum and under concentration were achieved with III-V based devices, reaching efficiencies as high as 47.1 %.[30–32]

To test the assumptions made in Equation (3), a $Ga_{0.91}In_{0.09}As_{0.83}P_{0.17}$ solar cell with almost ideal photovoltaic properties in terms of diode behavior and transport losses as derived from a one-diode model fit to a measured current-voltage characteristic was studied (cf. SI). Figure 1 displays the current voltage characteristic of this III-V semiconductor device and the simultaneously recorded PL(V) curve. The remarkable overlap of these measurements highlights the validity of the above outlined theoretical considerations. .

By applying a baseline correction considering the reflectance of the sample, at short circuit ($J(0) = J_{sc}$), the charge extraction efficiency $J_{sc}/J_{gen}|_{PL}$ is assessed to be 0.87 from the PPI measurements. To validate this result, as the internal quantum efficiency (IQE) was determined by a direct measurement of the external quantum efficiency (EQE) and the reflectance of the device at a wavelength of 620 nm, close to the excitation wavelength of the PPI setup of 623 nm. The IQE value of 0.98 (cf. SI) is in good agreement to $J_{sc}/J_{gen}|_{PL}$. The remaining difference of around 10 % can likely be attributed to the propagation of measurement uncertainties from all the methods that were involved.

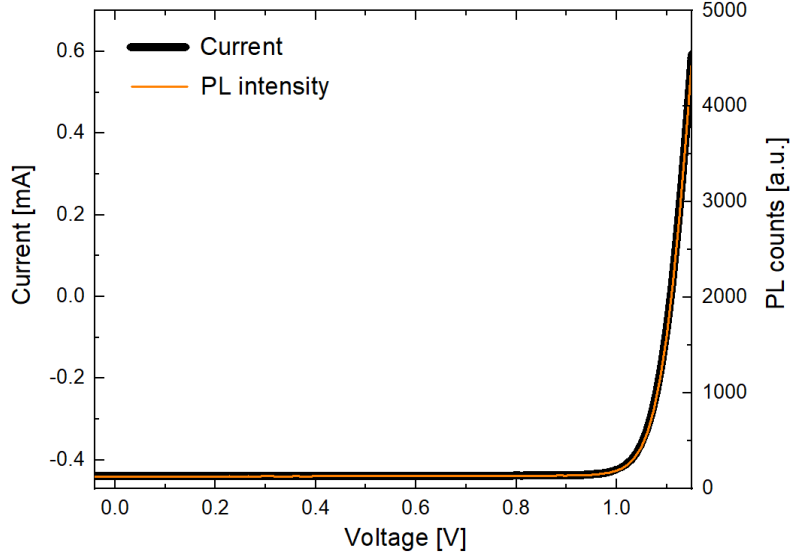

**Figure 1.** Current-voltage characteristic of a highly efficient (18.7 % power conversion efficiency) $Ga_{0.91}In_{0.09}As_{0.83}P_{0.17}$ solar cell (thick black line), compared with the simultaneously recorded PL(*V*) characteristic (thin orange line).

**Perovskite-based photovoltaic device**

Monocrystalline, wafer-based PV devices like III-V solar cells display very homogeneous photoluminescence features (cf. Figure S5). To assess the full potential of microscopically resolved charge extraction imaging by the PPI approach, measurements were carried out on a high-efficiency perovskite solar cell with a layer structure of glass / fluorine-doped tin oxide (FTO) / compact titanium dioxide ($TiO_2$) / mesoporous $TiO_2$ / perovskite / Spiro-OMeTAD / Au. Stabilized measurements of the *J-V* parameters under AM 1.5g illumination yielded a stabilized PCE of 21.2 %, a stabilized $J_{sc}$ of 26.4 mA/cm², and a stabilized $V_{oc}$ of 1.03 V (cf. Figure S4 in the SI).

Despite the high performance, the microscopic PL-image of the device recorded at open circuit conditions (PL($V_{oc}$)) shown in Figure 2a displays a highly inhomogeneous pattern of the active area. The image reveals morphological patterns that are typically assigned to perovskite films formed by spin coating such as stripes and dots. In the region of highest PL($V_{oc}$) such as spot (i), the PL intensity exceeds that of the surrounding area by more than a factor of five. Figure 2b displays the image of the local short circuit current assessed by PPI measurements in steady state, i.e., after stabilizing the devices for longer than 60 s. As can be seen in the $J_{sc}/J_{gen}|_{PL}$ image, it can be misleading to directly conclude from the PL($V_{oc}$) image on the local device performance. In the $J_{sc}/J_{gen}|_{PL}$ representation, the active area is relatively more homogeneous and will reveal more meaningful information on the aforementioned morphological patterns. Concerning regions of high PL($V_{oc}$), the PPI method allows to distinguish between two distinct features: feature (i) displays a spot with very low $J_{sc}/J_{gen}|_{PL}$, i.e., the PL intensity is not quenched when switching from open to short circuit. While the high PL($V_{oc}$) is a clear indicator for the presence of high-quality perovskite crystals with low non-radiative recombination, the low $J_{sc}/J_{gen}|_{PL}$ suggests a poor electrical connection of the perovskite to the charge extraction layers. In contrast, feature (ii) displays both high PL($V_{oc}$) and high $J_{sc}/J_{gen}|_{PL}$, due to a benign electrical coupling. A detailed discussion on experimental limitations of the approach and calibration strategies is presented in the SI.

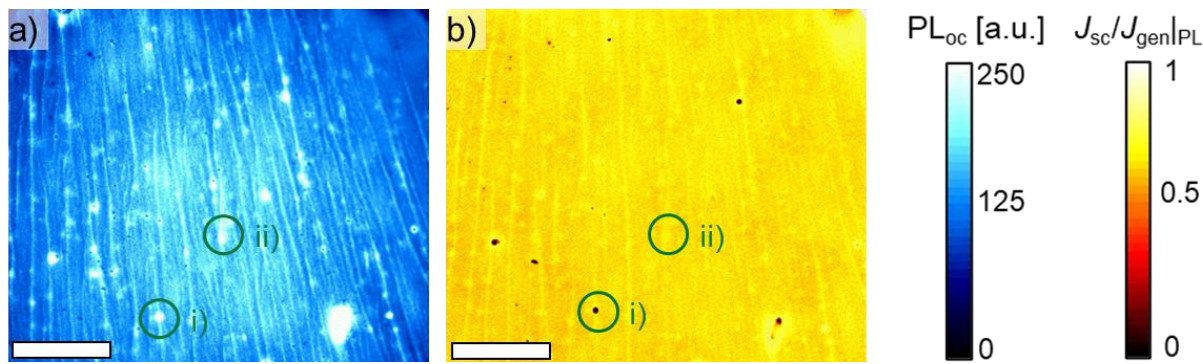

**Figure 2.** a) PL($V_{oc}$) and (b) $J_{sc}/J_{gen}|_{PL}$ determined from two PL images at open and short circuit of a high performing (>20%) perovskite solar cell. The device was illuminated with a red LED (632 nm). The scale bar corresponds to 400 μm in both images.

## *J(V)* imaging

Figure 3a shows the microscopic PL($V_{oc}$) image of another perovskite solar cell, displaying similar patterns as the sample discussed above. The patterns were grouped according to the pattern type and PL intensity in the PL($V_{oc}$) image: dots (A) and stripe patterns (B) with lower PL($V_{oc}$) as well as stripes with higher PL($V_{oc}$) (C) (cf. magnified image in Figure 3b) . With the PPI method, we can retrieve the local $J(V)/J_{gen}|_{PL}$ curve for each pixel of the microscope image. Figure 3c shows the corresponding $J(V)/J_{gen}|_{PL}$ curve for the features A, B, and C. One can see that the similarities from the PL($V_{oc}$) images are also found in the $J(V)/J_{gen}|_{PL}$ representation, where the $J_{sc}$ is highest for features C, lower for B and lowest for A. This demonstrates the potential to directly record the local $J(V)/J_{gen}|_{PL}$ image with microscopic resolution.

At this point, it is important to note that the method implicitly assumes that the $V_{oc}$ is equal over the entire cell area. This assumption is only valid if the absorber is well coupled to the outer electrodes and if the sheet resistivity of the electrodes is low. Finally, it is important to take the unique transient behavior of perovskite material into account for any luminescence measurement.[24] For PSC, reaching a stabilized state after changes in electrical bias, illumination, or even atmospheric conditions can require minutes to hours.[33–35] Moreover, non-reversible degradation processes can take place in these timeframes and under these conditions.[36,37] A discussion of the transient behavior can be found in the SI.

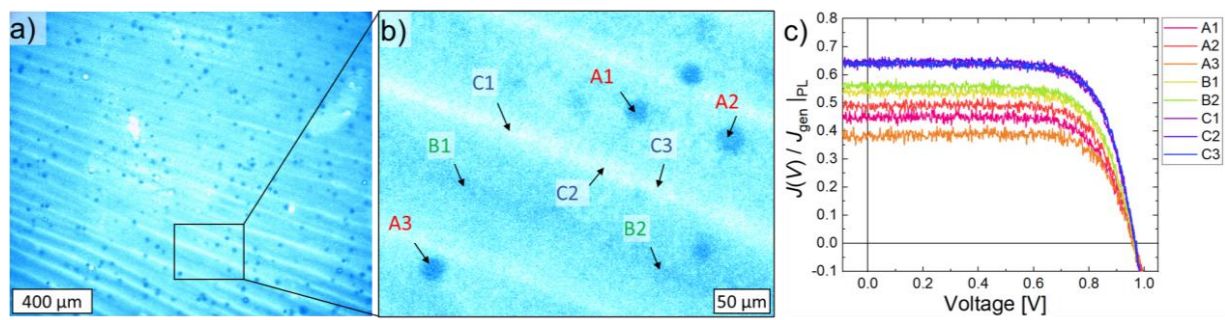

**Figure 3.** a) PL($V_{oc}$) intensity map of a sample in open circuit and b) closeup of a selected area. c) Local $J(V)/J_{gen}|_{PL}$ curve of the spots highlighted in (b) for a slow reverse voltage scan (20 mV/s). The sample was illuminated with a 632 nm LED light.

## The influence of interfaces on charge extraction

To investigate the effect of the different PSC layers on local extraction of photogenerated charge carriers, a range of spin-coated perovskite cells with an n-i-p structure were studied for which the optimal layer configuration of glass / TCO / electron-extraction layer / perovskite / hole-extraction layer / Au was intentionally altered. Figure 4 shows the microscopic images of PL($V_{oc}$) and $J_{sc}/J_{gen}|_{PL}$, respectively, for a solar cell with reduced perovskite layer thickness (a, b), a device without electron-extraction layer (EEL) (c, d), and one without hole-extraction layer (HEL) (e, f). The $J_{sc}/J_{gen}|_{PL}$ averaged over the entire image is 0.71, 0.24, and 0.11, respectively.

The images reveal a range of interesting features and general observations: in Figure 4a and b, stripe or wave patterns can be recognized that are most probably caused by the spin-coating process. For these patterns, high PL($V_{oc}$) correlates with high $J_{sc}/J_{gen}|_{PL}$. Comparing the two images furthermore demonstrates that by only employing the PL($V_{oc}$), some features remain concealed which become visible in the $J_{sc}/J_{gen}|_{PL}$-representation. In this sense, features (iii) and (iv) in the bottom of the image are most striking for which the low $J_{sc}/J_{gen}|_{PL}$ indicates poor charge extraction. In feature (iii), also the PL($V_{oc}$) is low, indicating that the perovskite is severely degraded (or absent) at this spot. In contrast, for feature (iv), the PL($V_{oc}$) in the same range as for the surroundings which shows that photovoltaic active material is present here. However, the $J_{sc}/J_{gen}|_{PL}$ image reveals that the photogenerated charges are not extracted.

Looking at the EEL-free devices in Figure 4c and d, we observe an absence of the wave pattern. Hence, this pattern is most probably induced by the spin-coating process of the EEL. Two additional features appear here. Similar to spot (v), feature (vi) has a high PL($V_{oc}$) but a low $J_{sc}/J_{gen}|_{PL}$. Inversely, there are many small spots, as represented by three circles for features (iv) that are prominent due to a high $J_{sc}/J_{gen}|_{PL}$. In the PL($V_{oc}$) image, they can, however, not be distinguished from the surrounding regions.

The HEL-free devices again do show the wave-pattern (Figure 4e, f). The PL($V_{oc}$) is much lower than for the other samples (note that the color scale was divided by a factor of 10). Also, the $J_{sc}/J_{gen}|_{PL}$ is low throughout most of the observed area. There are a few small dots as represented by feature (v) that display both relatively high PL($V_{oc}$) and high $J_{sc}/J_{gen}|_{PL}$. In contrast, similar to features (iii) and (iv), feature (viii) shows high PL($V_{oc}$) but low $J_{sc}/J_{gen}|_{PL}$.

Further assessments of the effect of the metal back electrode and a comparison of *I-V* parameters with PL data can be found in the SI.

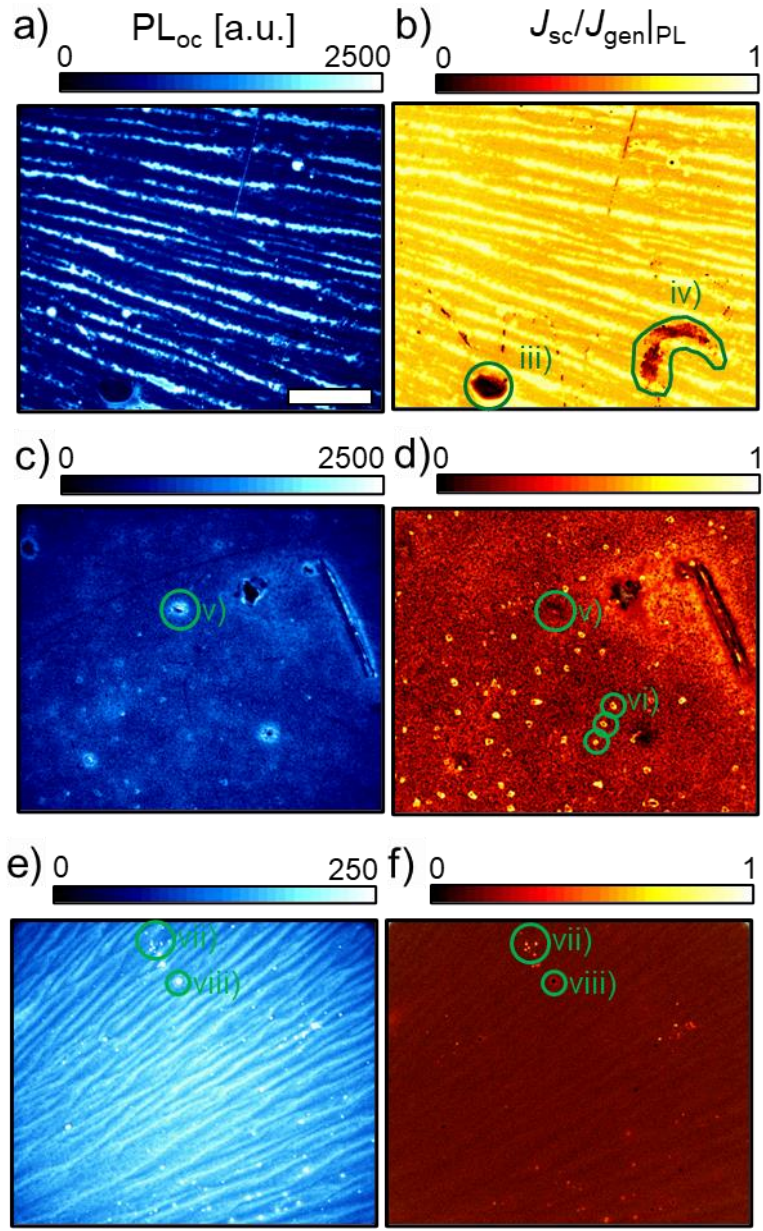

**Figure 4.** PL at open circuit (PL($V_{oc}$), blue images) and normalized short circuit photocurrent density ($J_{sc}/J_{gen}|_{PL}$, red images) imaging analysis. Three different spin-coated n-i-p PSC are shown, respectively: A cell with reduced perovskite layer thickness (a, b) as well as an EEL-layer free (c, d) and an HEL-layer free device (e, f). Note that the PL($V_{oc}$) intensity of images a) and c) ranges up to 2500 counts whereas image e) ranges only up to 250 counts. The samples were illuminated with a 632 nm LED light. The scale bar represents 400 μm for all images.

**Identification of charge extraction loss mechanisms**

The measurements presented above demonstrate that the $J(V)/J_{gen}|_{PL}$ approach is a simple and powerful tool to analyze the local performance of perovskite solar cells with microscopic resolution. As the $J_{sc}/J_{gen}|_{PL}$ representation also contains information on the local PL($V_{oc}$) and PL($V$=0), we can use this method to unravel interfacial charge extraction and recombination mechanisms that would not be possible using methods like LBIC alone. This means that we are not only able to detect regions of low photocurrent but can also estimate why the current is low. Such an investigation is especially significant for the characterization of solution-processed solar cells like PSC where a key challenge is the establishment of good electrical coupling between the perovskite crystals and the charge extraction layers.[38]

Figure 5a illustrates how $J_{sc}/J_{gen}|_{PL}$ can be span by the corresponding PL($V_{oc}$) and PL($V=0$) intensities. A high $J_{sc}/J_{gen}|_{PL}$ is achieved if PL is high at open circuit and zero at short circuit. On this map, we can now classify features of the above studied samples within a set of extreme cases.

1. In the **ideal case**, $J_{sc}/J_{gen}|_{PL}$ approaches unity. This means that the non-radiative recombination is minimized, and radiative recombination is maximized. Practically, this situation is reached if there is full coverage of a high-quality perovskite crystal layer (corresponding to a maximal PL($V_{oc}$) at open circuit), while the photogenerated charge carriers are ideally extracted at short circuit. This is represented by the white area at the ordinate of the graph in Figure 5a and by situation (1) in Figure 5b where a schematic cross-section of a PSC is displayed. Features (ii), (vi), and (vii) approach this ideal situation.

A $J_{sc}/J_{gen}|_{PL}$ of zero occurs for two worst-case scenarios:

2. There is **no presence of a functional perovskite layer**, as for the case of degraded perovskite or even complete absence of this layer, as depicted by situation (2) Figure 5b. This situation can be identified if PL($V_{oc}$) is also zero, as represented by the origin in the graph of Figure 5a). Such a situation can be attributed to feature (iii)

3. A perovskite layer is present, but there is **no efficient charge extraction** of the photogenerated charge carriers to the outer terminals. In the most dramatic case, this occurs in the absence of a back electrode, as illustrated by case (3) in Figure 5b. In this case, the PL is not changed at short circuit and thus PL($V_{oc}$) = PL($V=0$), represented by the black area at the bisector in Figure 5a. Due to a lower surface recombination, in practice the PL($V_{oc}$) is likely higher in these regions in comparison to those with good connection to contact layers. Features (v) and (viii) are representative for this situation.

4. Finally, there is also the possibility of **poor charge selectivity**. This can be the case for an absence of charge selective layers such that the back or front electrode is directly in contact with the perovskite absorber (case (4, 4') in Figure 5b). Another possibility is a high surface recombination/ poor selectivity of the EEL or HEL. In this case, the PL($V=0$) can be zero, but due to high surface recombination, PL($V_{oc}$) is also low. The entire active area of the devices without EEL or HEL displayed in Figure 4d and f are representative for this situation.

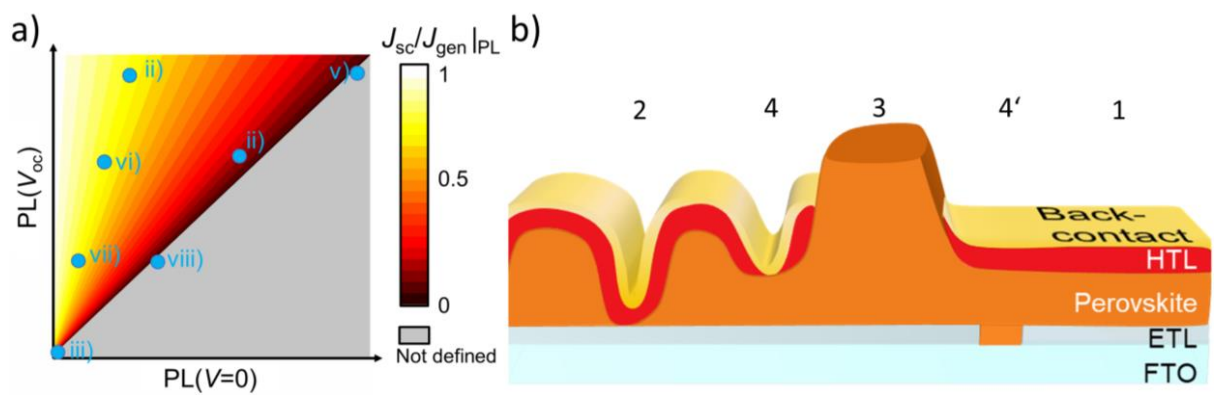

**Figure 5.** a) Evolution of the charge extraction coefficient respective to short circuit, $J_{sc}/J_{gen}|_{PL}$, for different values of PL intensity at open and short circuit. The blue dots correspond the features from Figures 2 and 4. b) Schematic illustration of the cross section of a n-i-p perovskite solar cell, displaying four characteristic cases for (non) ideal charge extraction.

## Conclusion

The photocurrent-voltage ($J(V)$) curve is the key characteristic for assessing photovoltaic devices. However, $J(V)$ measurements usually do not contain any spatial information, which are essential for a deeper understanding and further improvements of the devices. A range of methods for the local assessment of the photocurrent have been developed, especially for silicon and III-V solar cells. Yet, none of them has so far been able to offer a combination of the possibility to assess the local $J(V)$ for all bias voltages between 0 V and $V_{oc}$ while at the same time yielding fast image acquisition at microscopical resolution.

With the PPI approach, we introduced a straightforward method for a time-resolved assessment of the local photovoltaic $J(V)$ curve of perovskite solar cells by electrical-bias dependent luminescence microscopic imaging. We derived from basic principles how the measurement of the difference in PL($V$) to the PL($V_{oc}$) at open circuit can be related to the local $J(V)$. The validity of the approach was experimentally demonstrated by applying it to a close-to-ideal III-V solar cell. Challenges and opportunities of the approach were systematically studied and discussed. By the investigation of a high performing (>20 %) perovskite solar cell, we showed that the PPI approach allows the real-time assessment of the microscopically resolved local assessment of the $J(V)$ curve, the steady-state $J_{sc}$ as well as transient charge extraction behavior. Furthermore, we demonstrated that the technique also reveals information to assess the local charge extraction efficiency and interfacial recombination mechanisms. It is therefore a valuable tool for the understanding of the electrical coupling of the perovskite to the charge extraction layers. This work shows that, combined with potentiostatic measurements, microscopic luminescence imaging can be a powerful tool for the assessment of performance losses and the improvement of solar cells. The introduced technique can make a significant impact on the understanding and improvement of perovskite and other solar cell technologies as it enables relating morphological artifacts to device performance at a microscopic resolution.

## Methods

The photoluminescence intensity was measured in reflectance with an optical microscope at 10x magnification. A red and blue high-power LED (Thorlabs Solis) with a peak intensity of 632 and 405 nm, respectively, was used to illuminate the solar cells at a light intensity equivalent to one sun as determined by measuring a reference silicon PV cell and calculating the photon flux from the EQE spectrum. On the detector side, the PL light was focussed onto a high resolution sCMOS camera (Andor Zyla 5.5) with an optical high-pass filter (cut-off at 760 nm). To apply an electrical bias and to simultaneously measure solar cell currents and voltages, a potentiostat (Ivium CompactStat) was used. After changing the electrical bias, the PL was stabilized for at least 60 seconds before recording the image of perovskite devices.

Stabilized current-voltage parameters such as PCE, $J_{sc}$ and $V_{oc}$ were determined by under a class A solar simulator whose intensity was calibrated with a silicon reference cell, where for perovskite devices 'stabilized' refers to the value measured after 100 s of measurement. For the determination of the stabilized PCE, the device current was recorded under a fixed bias voltage close to the suspected maximum power point.

Processing parameters of solar cells are outlined in the SI.

## Acknowledgements

L.W., J.P.H., S.M., and U.W. acknowledge the funding from the European Union's Horizon 2020 research and innovation program under Grant Agreement No.763989 (APOLO). This publication reflects only the author's views and the European Union is not liable for any use that may be made of the information contained therein. L.W., M.E., D.B., S.Z., S.M. and A.H. acknowledge funding within the project UNIQUE, supported under the umbrella of SOLAR-ERA.NET_Cofund by ANR, PtJ,

MIUR, MINECOAEI, SWEA. SOLAR-ERA.NET is supported by the European Commission within the EU Framework Programme for Research and Innovation HORIZON 2020 (Cofund ERANET Action, No. 691664). L.W. and D.B. acknowledge scholarship support by the German Federal Environmental Foundation (DBU). P. S. acknowledges his PhD scholarship from the Heinrich-Böll Foundation. S.Z. acknowledges scholarship support by the German Academic Exchange Service (DAAD). We thank Markus Glatthaar (Fraunhofer ISE) for valuable discussions.

## Author contributions

L.W. and A.H. developed the idea of the PPI approach, derived the theory and conceived the initial measurement. The concept was intensively discussed and the measurement method was refined by the significant contribution of L.W., J.P.H., P.S., M.E., D.B., S.Z., S.M. U.W., A.H. and S.G.. L.W. wrote the manuscript, developed, and carried out PPI measurements and data analysis on III-V devices together with P.S. and on PSC, together with J.P.H., M.E., D.B., and S.Z. Perovskite samples were provided by and data on PSC was discussed with at Y.L, S.M.Z, and M.G.. III-V samples were fabricated and further characterized by P.S. All authors discussed the results and revised the manuscript

# Supporting Information

**Revealing Fundamentals of Charge Extraction in Photovoltaic Devices Through Potentiostatic Photoluminescence Imaging**

Lukas Wagner[1,2,*], Patrick Schygulla[1,2], Jan Philipp Herterich[1,3], Mohamed Elshamy[1], Dmitry Bogachuk[1,2], Salma Zouhair[1,4], Simone Mastroianni[1,3], Uli Würfel[1,3], Yuhang Liu[5,*], Shaik M. Zakeeruddin[5], Michael Grätzel[5], Andreas Hinsch[1,*], Stefan W. Glunz[1,2]

[1] Fraunhofer Institute for Solar Energy Systems ISE, Heidenhofstraße 2, 79110 Freiburg, Germany.
Email: lukas.wagner@ise.fraunhofer.de, andreas.hinsch@ise.fraunhofer.de

[2] Laboratory for Photovoltaic Energy Conversion, University of Freiburg, Emmy-Noether-Straße 2, 79110 Freiburg, Germany.

[3] Freiburg Materials Research Center FMF, University of Freiburg, Stefan-Meier-Straße 21, 79104 Freiburg, Germany.

[4] Abdelmalek Essaadi University, FSTT, Thin Films & Nanomaterials Lab, 90000 Tangier, Morocco.

[5] Laboratory of Photonics and Interfaces (LPI), Department of Chemistry and Chemical Engineering, École Polytechnique Fédérale de Lausanne, Lausanne CH-1015, Switzerland.
E-mail: yuhang.liu@epfl.ch

# Contents

# A. Discussion of established imaging and mapping techniques

In solar cell research, mainly two approaches are followed for spatial resolved characterization of the active cell area, namely scanning-based "mapping" and camera-based "imaging" techniques. A range of approaches to determine the local photocurrent have been established for wafer-based solar cells such as GaAs or silicon photovoltaics (PV) and were also applied to emerging technologies like perovskite solar cells (PSC), namely light beam-induced local current (LBIC) mapping, lock-in thermography (LIT) imaging, as well as photoluminescence (PL) and electroluminescence (EL) based imaging techniques.[1,2] The opportunities and constraints of these characterization methods are briefly outlined in the following. An overview is presented in Table S1.

**LBIC**
The LBIC method allows recording photocurrent maps by locally illuminating the sample with a confined laser spot and measuring the induced local short-circuit photocurrent ($J_{sc}$) with high spatial resolution.[3–5] The acquisition time for one map at short circuit lies in the range of 30 minutes.

**LIT**
In dark lock-in thermography (DLIT) imaging, the local power dissipation of a non-illuminated sample is assessed by measuring the joule heating induced by an applied AC voltage via a lock-in infrared detector. The to assess the local $J(V)$ and local $J_{sc}$, several methods to fit equivalent circuit models to data acquired at different bias voltages have been proposed.[6,7] The weakness of all methods based on fitting data to an equivalent circuit model lies in the fact that, a range of datapoints are necessary for a good fit and, most importantly, the choice of the diode model has to be made *ex ante* and has a strong impact on the retrieved parameters.
Notably, for illuminated lock-in thermography (ILIT), an approach for Si-PV was developed based that allows assessing the local $J_{sc}$ without fitting to a diode model.[8] Yet, the approach is based on the assumption that the photocurrent at reverse bias does not differ significantly from the $J_{sc}$. For PSC, reverse bias can induce device degradation,[9,10] which might explain why, to the best of our knowledge, this ILIT approach was so far not reported for PSC. Both LIT methods have the drawback that the low signal-to-noise ration requires long data acquisition times per images in the range of 20 to 60 minutes.[3] Especially for coupled electric-ionic semiconductors like perovskites, this approach may lead to measurement artefacts due to sample degradation or transient ionic effects due to the AC bias voltage.

**Photoluminescence imaging**
Employing PL images, based on various equivalent circuit models, a range of refined approaches for the spatial determination of the dark saturation current and series resistance[11–16] as well as local short circuit current[17] have been proposed for Si-PV. For perovskite solar cells, PL imaging so far has mainly been applied to study layer uniformity, crystallinity and degradation in perovskite solar cells under open circuit (PL($V_{oc}$)).[3,18–23] Recently, PL based series resistance imaging methods have also been applied to PSC.[24,25]

**Electroluminescence imaging**
Electroluminescence imaging is based on the principle of inducing luminescence by injection of charge carriers via the application of a forward bias that is applied to the device in the dark. For photocurrent imaging, EL has, however, only the potential to assess the dark spatial $J(V)$ curves. Moreover, for perovskite PV meager signal-to-noise ratios are achieved even at forward bias above $V_{oc}$ and for acquisition times of several minutes. For PSC, this does not only limit the spatial resolution but can also falsify the result due to an integration over transient effects.[22]

**Table S1. Comparison of established photocurrent imaging / mapping techniques and the PPI method.**

|  | Data acquisition method | Data acquisition time per bias voltage (approximate) [min] | Applicable bias voltages | Steady state (spatially illumination) | Steady state (bias voltage) | Local $J(V)$ acquisition |
|---|---|---|---|---|---|---|
| **LBIC** | mapping | 30 | 0 V | no | yes | no* |
| **DLIT** | imaging | 20 – 60 | various | dark | yes | diode model fitting |
| **ILIT** | imaging | 20 – 60 | 0 V and < 0 V | yes | yes | no |
| **PL** | imaging | < 1 | various | yes | yes | diode model fitting |
| **EL** | imaging | 1 - 10 | > $V_{oc}$ | dark | yes | direct |
| **PPI** | imaging | < 1 | various | yes | yes | direct |

* For Si-PV, Carstensen et al. presented an approach to reconstruct the local photocurrent at different bias voltages from fitting diode-models to LBIC data, however not at microscopic resolution.[26]

## B. Theoretical relationship between the radiative and non-radiative recombination

### Ideality factor of recombination currents

If the studied sample can be described by the Shockley diode equation, we can express the recombination current density $J_{rec,i}$ of a specific recombination mechanism (i) by [27,28]

$$J_{rec,i}(V) = J_{rec,i,0} \left[ e^{\frac{\Delta E_f}{n_{id,i} k_B T}} - 1 \right], \tag{S1}$$

where $J_{rec,i,0}$ is a material specific parameter, $\Delta E_f$ is the quasi fermi-level splitting (QFLS), $k_B$ is the Boltzmann constant and $T$ is the temperature. $n_{id,i}$ is the ideality factor of the specific recombination mechanism.

If the voltage drop across the contact layers is negligible, we can approximate the QFLS by the voltage at the outer terminals $\Delta E_f = eV$. Deviations of this case are discussed in Section C.

For $\Delta E_f = eV$, we can simplify two cases: close to the short circuit case ($V \to 0V$)

$$J_{rec,i}(V) \to 0. \tag{S2a}$$

This means that in the ideal case, all charge carriers are extracted outside of the cell in short circuit.

If the applied voltage is larger than the thermal voltage ($V_T = k_B T/e \approx 25.8$ mV $\ll V$), which is the case except for voltages close to short circuit, the equation can be approximated by

$$J_{rec,i}(V) = J_{rec,i,0}\, e^{\frac{V}{n_{id,i} V_T}}, \tag{S2b}$$

on which we focus in the following consideration.

The ideality factor can be expressed by[29]

$$n_{id} = \frac{m}{\delta}. \tag{S3}$$

Herein, $m$ is a measure for the number of the charge carrier types ($n$ or $p$ for negative or positive charge carriers, respectively) that are involved in splitting the quasi Fermi levels: $np = n^m$ (or $p^m$). Herein $n = \Delta n + n_0$ and $p = \Delta p + p_0$, where $\Delta n, p$ being the (symmetrically) photoexcited charge carriers with $\Delta n = \Delta p$. $n_0$ is the background doping concentration. We can see that if the material is sufficiently doped ($n_0 \gg \Delta n$, $p_0 \ll \Delta p$) then $n \gg p$ (or vice versa) and hence $m$ approaches one.

If the photogenerated charge carriers greatly excess the background doping concentration ($\Delta n = \Delta p \gg p_0, n_0$, also known as high injection), then $n \approx p$ and $m \to 2$. This is most probably the case for typical perovskite photoabsorbers which, in contrast to Si-PV, have been reported to behave like intrinsic or low-doped semiconductors. [30–33]

The ideality factor is also influenced by the recombination reaction order $\delta$. In open-circuit conditions, the reaction order describes how the effect of the charge carrier density $n$ on the specific recombination

$$R_i = k_i n^\delta, \quad (S4)$$

with $k_i$ being the recombination specific rate constant. Hence, $\delta$ can be imagined as the number of charge carriers that are involved in a recombination process (at the time when the process happens).

For the $n \approx p$, $m \to 2$ case, there are two recombination processes with clearly defined recombination orders: radiative and Auger recombination (the remaining Shockley-Read-Hall (SRH) recombination is discussed further below). For radiative or band-to-band recombination, two charge carriers are involved as a free electron recombines with a free hole, thereby emitting a photon. We can write

$$R_{rad} = k_{rad} np, \quad (S5)$$

and hence $\delta = 2$. For the recombination current this means that $n_{id,rad} = 1$ and

$$J_{rad}(V) = J_{rad,0} e^{\frac{V}{V_T}}. \quad (S6)$$

For Auger recombination, the recombination of an electron and a hole leads to the excitation of a third charge carrier, hence $\delta = 3$ and $n_{id,Auger} = 2/3$. In contrast to Si-PV, PSC display low Auger recombination rate constants $k_3$,[33] which means that under normal illumination intensities around 1 sun, the contribution of Auger recombination to non-radiative recombination processes can be neglected.

The remaining contribution to non-radiative (SRH) recombination for PSC is still under debate and a range of different recombination processes with different specific ideality factors have been reported.[28,34] According to the considerations above $n_{id,n.r.}$ needs to lie between 1 and 2. If the ideality factor if the entire $I$-$V$ curve close to one then it is reasonable to assume that $J_{n.r.}$ is either negligible or that $n_{id,n.r.} \approx 1$. In this case, $J_{n.r.}$ is linearly proportional to $J_{rad}$ as expressed by $k$ = const. in Equation (2).

**Generalized form of Equation (5)**

In the following, we consider the general case where

$$J_{n.r.}(V) = J_{n.r.,0} e^{\frac{V}{n_{id,nr} V_T}}. \quad (S7)$$

Now, Equation (2) can be written as

$$J_{n.r.}(V) = k' \cdot J_{rad}(V)^{\frac{1}{n_{id,nr}}}, \quad (S8)$$

with $k'$ = const.

Then, Equation (1) yields

$$J(V) = J_{gen} - J_{rad}(V) - J_{n.r.}(V) = J_{gen} - J_{rad}(V) - k' \cdot J_{rad}(V)^{\frac{1}{n_{id,nr}}} \quad (S9)$$

Hence, if the relationship between $J_{rad}$ and $J_{n.r.}$ is unknown, then

$$S_{PL}(V_{oc}) - S_{PL}(V) = \frac{c}{e}[J_{rad}(V_{oc}) - J_{rad}(V)] \tag{S10}$$

$$= \frac{c}{e}[J_{gen} - J_{n.r.}(V_{oc}) - J_{gen} + J(V) + J_{n.r.}(V)]$$

$$= \frac{c}{e}[J(V) - \Delta J_{n.r.}(V)],$$

with $\Delta J_{n.r.}(V) = J_{n.r.}(V_{oc}) - J_{n.r.}(V)$.

The normalized difference between the PL signal at two voltages can now be expressed in the general form of

$$\frac{S_{PL}(V_{oc}) - S_{PL}(V)}{S_{PL}(V_{oc})} = \frac{J(V) - \Delta J_{n.r.}(V)}{J_{gen} - J_{n.r.}(V_{oc})} =: c_{ex}(V). \tag{S11}$$

In conclusion, for the case of $n_{id,n.r.} \neq 1$ the approach does not yield an exact assessment of the local photocurrent since the local non-radiative recombination currents at $V_{oc}$ are unknown. Yet, the approach can be still employed as a figure of merit for local charge extraction, as expressed by a charge extraction coefficient $c_{ex}(V)$.

Finally, we point out that by comparing the $c_{ex}(V)$ image with a complementary photocurrent imaging/mapping method the approach can be used to qualitatively assess the spatially resolved non-radiative ideality factor or internal voltage loss which gives instructive insights in the internal loss mechanisms that induce recombination and hamper charge extraction.

## C. Accounting for internal voltage and current losses

The condition $\Delta E_f = eV$ in Equation S1 may not be fulfilled die to an internal voltage loss. This can occur inside the photoactive layer if there is a diffusion limited current which creates an implied voltage inside the device.[15] The voltage can also be lost across contact layers due to due to contact resistance at the interfaces with CELs or due to resistance of the CEL itself.[35] Finally, photogenerated charge carriers may be lost due to shunt resistance, e.g., via pinholes that allow electrical contact between the CELs. In a first order approximation, such processes can be accounted for if we consider local series and shunt resistances, $R_s$ and $R_{sh}$, respectively. Now, we can distinguish between the current densities and voltage measured at the outer solar cell cables, $J(V)$ and $V$, respectively, and the internal current density and voltage, $J_{int}(V_{int})$ and $V_{int}$, respectively, as depicted in Figure S5. For the sake of readability, in the following ‚R' represents the resistances divided by the considered area.
Figure S5 shows that

$$J(V) = J_{int}(V_{int}) - \frac{V_{int}}{R_{sh}} = J_{int}(V_{int}) - \frac{J(V) \cdot R_s + V}{R_{sh}}. \tag{S12}$$

If the solar cell follows the Shockley equation, as discussed above we can write

$$J_{int}(V_{int}) = J_{gen} - J_{rad,0}\left[e^{\frac{V+J(V)\cdot R_s}{n_{id,rad}V_T}} - 1\right] - J_{n.r.,0}\left[e^{\frac{V+J(V)\cdot R_s}{n_{id,nr}V_T}} - 1\right]. \tag{S13}$$

This yields

$$J(V) = J_{gen} - J_{rad,0}\left[e^{\frac{V+J(V)\cdot R_s}{n_{id,rad}V_T}} - 1\right] - J_{n.r.,0}\left[e^{\frac{V+J(V)\cdot R_s}{n_{id,nr}V_T}} - 1\right] - \frac{V+J(V)\cdot R_s}{R_{sh}}. \tag{S14}$$

For Equation S11, this implies

$$\frac{S_{PL}(V_{oc}) - S_{PL}(V)}{S_{PL}(V_{oc})} = \frac{J_{rad,0}\left[e^{\frac{V_{oc}+J(V_{oc})\cdot R_s}{n_{id,rad}V_T}} - 1\right] - J_{rad,0}\left[e^{\frac{V+J(V)\cdot R_s}{n_{id,rad}V_T}} - 1\right]}{J_{rad,0}\left[e^{\frac{V_{oc}+J(V_{oc})\cdot R_s}{n_{id,rad}V_T}} - 1\right]} \quad (S15a)$$

or

$$\frac{S_{PL}(V_{oc}) - S_{PL}(V)}{S_{PL}(V_{oc})} = \frac{J(V) - \frac{V + J(V)\cdot R_s}{R_{sh}} - \Delta J_{n.r.}(V + J(V)\cdot R_s)}{J_{gen} - J_{n.r.}(V_{oc} + J(V)\cdot R_s)}. \quad (S15b)$$

In other words, the working point of the radiative recombination current, i.e., the PL signal, is influenced by the series resistance. Moreover, the shunt resistance induces an additional recombination path for the (internal) photocurrent.

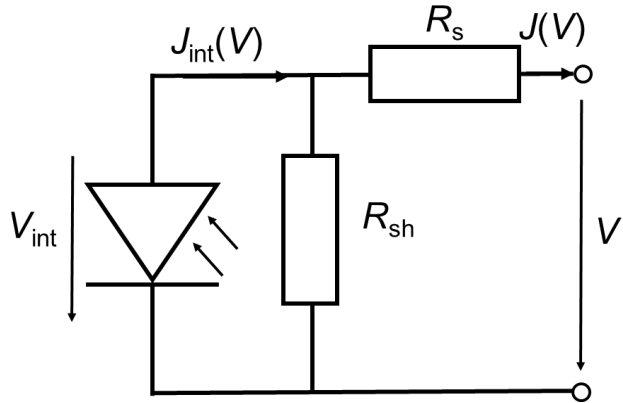

**Figure S5. Equivalent circuit of a solar cell spot with shunt and series resistances, $R_s$ and $R_{sh}$, respectively.**

## D. Epitaxial Growth and Characterization of a III-V solar cell

The GaInAsP solar cell was grown using metalorganic vapor phase epitaxy on a commercial AIXTRON AIX2800G4-TM reactor. Trimethylgallium and trimethylindium were used as precursors for the group-III elements, and arsine and phosphine for the group-V elements. Silane was used for n-type and dimethylzinc for p-type doping. Standard growth temperatures and V/III ratios were used. The cell structure is depicted in Figure S2. It consists of a thick n-type $Ga_{0.91}In_{0.09}As_{0.83}P_{0.17}$ absorber above a heterojunction interface to a higher band gap $Al_{0.3}Ga_{0.7}As$ layer. This rear-heterojunction cell architecture allows exploiting the higher carrier lifetimes in n-type $Ga_{0.91}In_{0.09}As_{0.83}P_{0.17}$ and reducing recombination in the space charge region.[36] More details on the growth process are reported elsewhere [37].

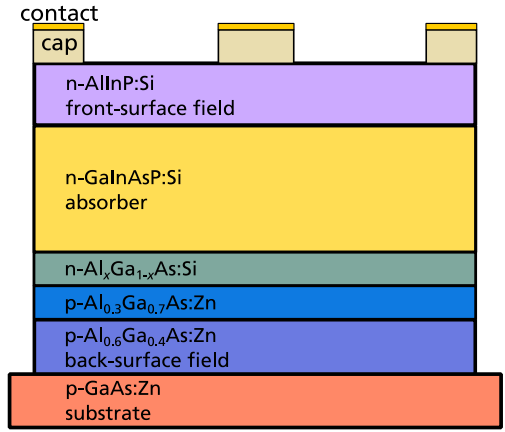

**Figure S6. Schematic solar cell layer structure of the epitaxially grown GaInAsP solar cell.**

From the epitaxial layer structure two types of solar cells were processed on the same wafer: larger devices with a mesa area of 4 cm² and smaller devices with a mesa area of 0.06 cm². The current-voltage characteristics of one of the large solar cells was measured under calibrated conditions with an AM1.5g solar spectrum in the Calibration Laboratory of Fraunhofer ISE (CalLab). The *I(V)*-curve was modelled using a one-diode equation [38]:

$$J(V) = J_0 \left( \exp\left(\frac{qV - R_s J(V)}{nkT}\right) - 1 \right) + \frac{qV - R_s J(V)}{R_p} - J_{ph}. \tag{S16}$$

The measured and modelled curve are both shown in Figure S7. The cell exhibits a diode ideality factor of 1.04, a dark recombination current of $8.23 \times 10^{-21}$ mA/cm², a very high shunt resistance of 9.19 kΩ cm², and a very low series resistance of 0.603 Ω cm². In order to confirm the low series resistance a suns-$V_{oc}$ measurement was conducted on a WCT-120 Sinton Instruments device.[39] Thanks to its currentless measuring mode this method allows to obtain a relative *I(V)*-curve without series resistance effects that can be scaled to the short-circuit current. In Figure S7 the standard *I(V)*-curve is compared to the suns-$V_{oc}$ *IV*-curve. Both curves coincide to a high degree. Around the open-circuit point the suns-$V_{oc}$-curve is even 10 meV below the calibrated curve which is likely due to the slightly different spectrum used for the suns-$V_{oc}$-measurement.

The PPI measurements, which employ an excitation spot much smaller than the area of the large cell, were performed on the small device in order to achieve an excess carrier density similar to AM1.5g. For the reported PPI and PL data, detector counts of a device region without metallization grid was selected. The external quantum efficiency (EQE) was measured on a large cell. Its absolute height was normalized so that the integrated photocurrent matches the short-circuit current density of the calibrated AM1.5g *I(V)*-curve (correction factor 1.07). The internal quantum efficiency $IQE =$

$EQE/(1-R)$ was then calculated from the EQE by including the reflectance $R(620 \text{ nm}) = 0.286$ (cf. Figure S8). The procedure to measure EQE and PPI on two different cells was necessary because of the difficulty of an accurate EQE measurement on a small solar cell. Considering the effects of shadowing due to a different metal grid on these two devices (correction factor 1.03), and given the fact that both cells have exactly the same epitaxial structure as they were processed on the same wafer, the EQE and PPI results can still be compared with each other. However, the resulting uncertainty of this comparison is expected to be high because of the high number of measuring techniques (EQE, reflectance, short-circuit current, PPI) involved that each add a little uncertainty to the calculation.

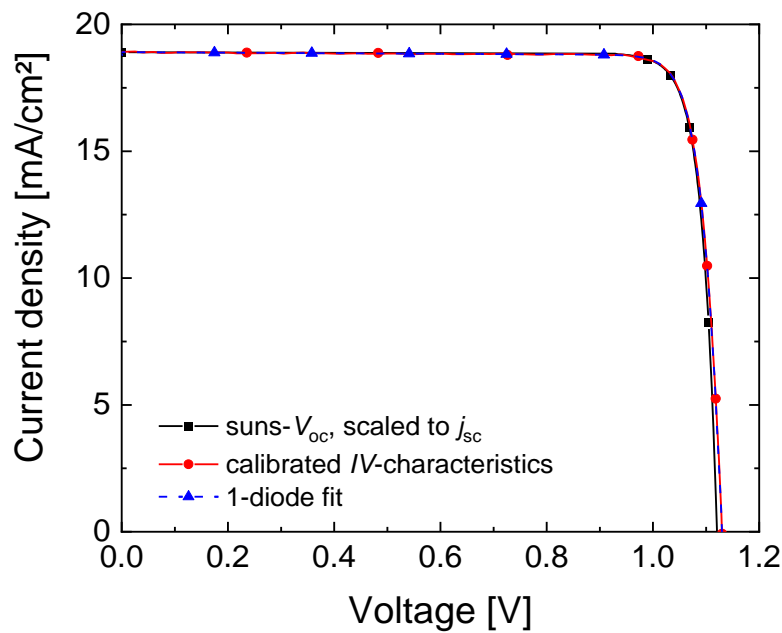

**Figure S7.** Comparison of a calibrated current-voltage characteristics under AM1.5g (red), suns-$V_{oc}$ characteristics scaled to the calibrated results (black), and a 1-diode model fit to the calibrated data (blue, dashed). The agreement between the curves shows the low series resistance, the high parallel resistance, and the ideality factor close to unity.

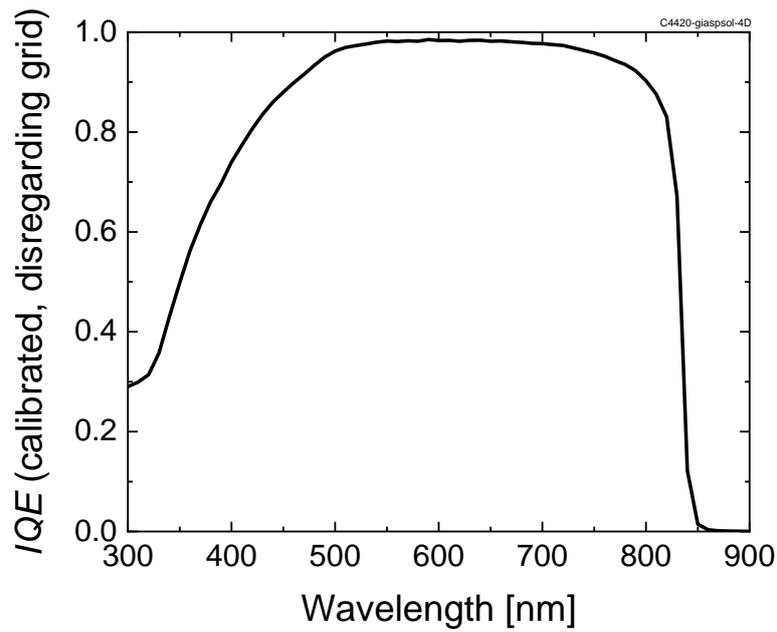

**Figure S8.** Internal quantum efficiency (*IQE*) of the 4 cm² GaInAsP solar cell as a function of wavelength. The curve was normalized to the calibrated AM1.5g short-circuit current density. The amount of grid shadowing was determined and corrected for so that a comparison to the PPI results, evaluated in a region of interest without metal grid, from the small solar cell becomes possible.

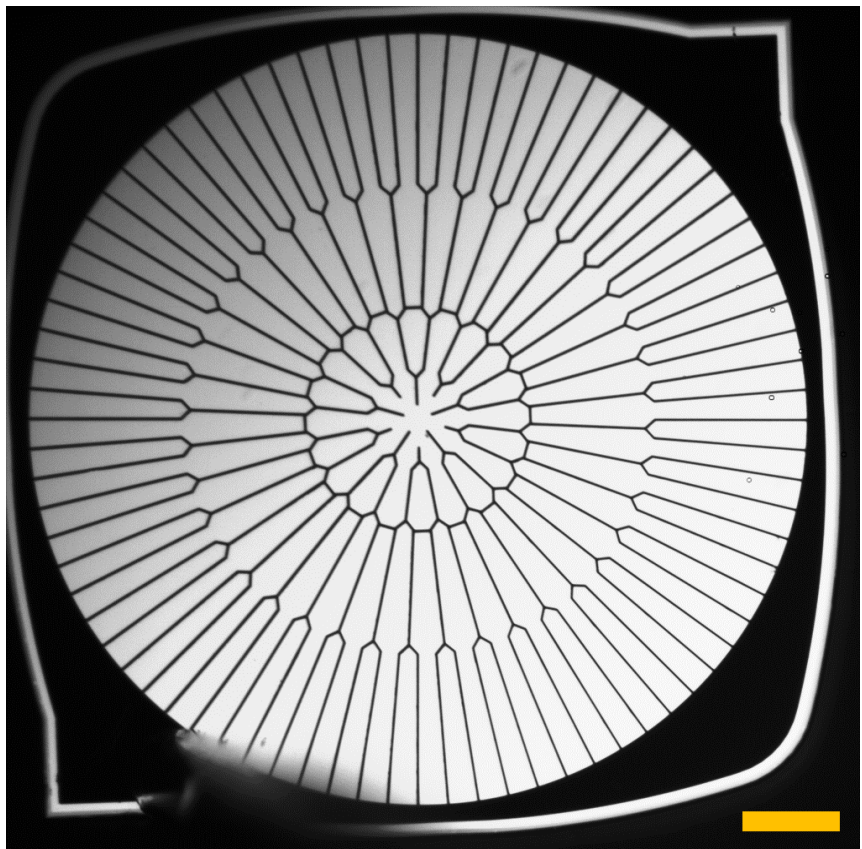

**Figure S9.** Photoluminescence microscopy image of the tested small GaInAsP cell in open circuit. The circular active area is covered with a front metallization grid. The yellow scale bar represents 300 µm.

# E. Perovskite solar cells fabrication

Materials: Lead iodide and cesium iodide are purchased from TCI co, ltd. Formamidinium iodide, methyl ammonium iodide and $N^2,N^2,N^{2'},N^{2'},N^7,N^7,N^{7'},N^{7'}$-octakis(4-methoxyphenyl)-9,9′-spirobi[9*H*-fluorene]-2,2′,7,7′-tetramine (Spiro-OMeTAD) are purchased from Dyesol. Ultra-dry dimethylformamide (DMF), ultra-dry dimethyl sulfoxide (DMSO) and ultra-dry chlorobenzene (CB) is purchased from Acros, dry isopropanol (IPA), 4-*tert*-butyl pyridine, lithium bistrifluorosulfonyl imide (LiTFSI), acetyl acetone, titanium diisopropoxide bis(acetylacetonate), 75 wt. % in isopropanol and borane tetrahydrofuran complex solution (1.0 M in THF) are purchased from Sigma-Aldrich. All the chemicals are used as received without further purification. Conductive glass, Fluorine-doped tin oxide (10 Ω/sq) is purchased from Nippon Sheet Glass, Titanium dioxide paste (30 NRD) is purchased from Dyesol.

Substrate preparation: Nippon Sheet Glass (NSG, 10 Ω/*sq*) was consecutively cleaned using 2% Hellmanex aqueous solution, deionized water, acetone and ethanol by sonicating for 20 min for each solvent. After drying with compressed air, UV-Ozone was applied for further cleaning. Compact $TiO_2$ is deposited on top of FTO using spray pyrolysis method: the substrates are preheated to 450 °C; a precursor solution of titanium diisopropoxide bis(acetylacetonate), 75 wt. % in isopropanol is diluted with dry ethanol with a volume ratio of 1:9 and addition 4% volume ratio of additional acetyl acetone. After spray pyrolysis, the FTO/$TiO_2$ substrate is allowed to heat at 450 °C for 30 min before cooling down to room temperature. Mesoscopic $TiO_2$ is applied by spincoating a diluted solution of 30 NR-D paste (mass ratio of paste:EtOH = 1:6) at 4000 rpm with acceleration of 2000 rpm/s. The as-prepared FTO/compact-$TiO_2$(c-$TiO_2$)/$TiO_2$ paste was then allowed to sinter at 450 °C for 1 h, yielding FTO/c-$TiO_2$/mesoscopic-$TiO_2$, which is then deposited with perovskite freshly.

Perovskite deposition: Pure 3D perovskite precursor solution is prepared by dissolving a mixture of lead iodide (736.5 mg, 1.60 mmol), formamidinium iodide (237.3 mg, 1.38 mmol), methylammonium iodide (9.5 mg, 0.06 mmol) and cesium iodide (15.6 mg, 0.06 mmol) in 1 mL mixed solution of DMF and DMSO (DMF (*v*):DMSO (*v*) = 4:1) under mild heating condition at ~70 °C to assist dissolving. The perovskite active layer is deposited using anti-solvent method, with chlorobenzene as anti-solvent. The perovskite precursor solution is deposited on the freshly-prepared FTO/c-$TiO_2$/m-$TiO_2$ substrate, a two-step spincoating method is applied. 1st step is proceeded at 1000 rpm with acceleration rate of 200 rpm/s for 10s. 2nd step is followed by 5000 rpm with acceleration rate of 2000 rpm/s for 20 s. 200 μL of CB is applied at the 10th second. After spin coating, the substrate is allowed to anneal at 110 °C for 40 min. The whole procedure is done in a nitrogen-filled glovebox.

Hole-transporting layer and gold back contact: Spiro-OMeTAD is selected as hole-transporting layer (HTL) material. Spiro-OMeTAD is dissolved in chlorobenzene with a concentration of 70 mM, which is doped by LiTFSI and 4-*tert*-butyl pyridine, and the molar ratios are 33% and 330%, respectively. The mixed Spiro-OMeTAD solution was spin-casted on the surface of the perovskite at 4000 rpm for 30 s. The acceleration is 2000 rpm/s. The gold electrode is thermally evaporated on the surface of the HTL with the shadow mask with an area of 5 mm * 5 mm. The thickness of gold electrode is 80 nm, and the evaporation speed is adjusted to 0.01 nm/s at the first 10 nm, and 0.08 nm/s for the rest of the procedure.

Photovoltaic performance measurements: Stabilized *I-V* parameters were measured under a class A xenon arc lamp solar simulator with a Keithley 2400 source meter. The light intensity was calibrated with a calibrated silicon cell. For the measurement of the stabilized PCE, the photocurrent was measured while the device was biased at a constant voltage close to maximum power point.

## F. Stabilized *I-V*-parameters of the high-efficient perovskite device

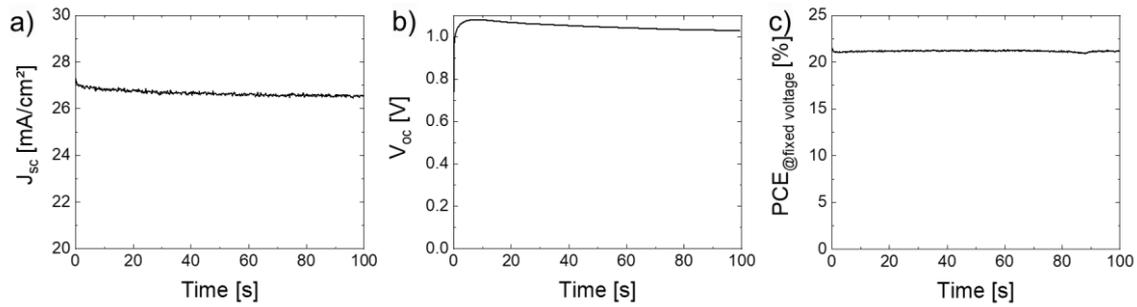

**Figure S10.** Stabilized $J_{sc}$, $V_{oc}$, and PCE (measured at a fixed voltage) of the high-efficient cell discussed in Figure 2. The device was measured under a class A solar simulator with a shadow mask of 0.09 cm².

## G. $J_{sc}/J_{gen}|_{PL}$ measurement of perovskite devices at different wavelengths

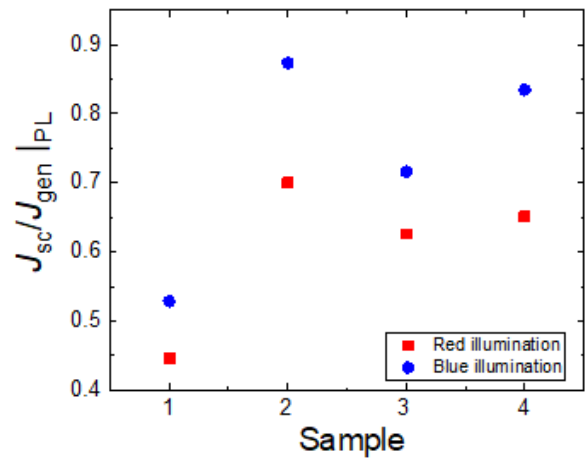

**Figure S11.** $J_{sc}/J_{gen}|_{PL}$ averaged over the entire microscope image for four different PSCs, probed under red (623 nm) and blue (405 nm) LED illumination.

## H. Calibration and error sources for qualitative $J_{sc}/J_{gen}|_{PL}$ imaging

**Uncertainty discussion**

Although the principle is straightforward, it is important to note that the significance of the here presented method is restricted to a range of fundamental and technical limitations which have to be assessed carefully. First, photon-reabsorption is not considered. Second, we cannot regard lateral diffusion currents that lead to an equilibration of charge densities from the 'high-performing' spots (which are slightly underestimated) and to the 'poorly performing' spots (which are slightly overestimated). Third, due to the light source of the measurement system, the depths of excitation wavelength can be different from an AM1.5g illumination (measurements under different excitation wavelengths are presented in Figure S11). Fourth, defining the steady state of the perovskite solar cell can be challenging and may vary from one sample to the other, as further discussed below. Finally, a detailed discussion of the effect of the diode ideality factor as well as internal (resistive) voltage and current losses can be found in Section B and C.

Moreover, we note from own experience that it is crucial to ensure homogeneous illumination of the entire active cell area. For samples with highly conductive electrodes, as present in the tested III-V and perovskite devices, illumination with a light spot smaller than the active area leads to a lateral distribution of charge carriers via the electrodes when the probe is biased in in open circuit. This induces radiative recombination in the non-illuminated regions and hence reduces the measured PL signal in the illuminated region. As this effect is less pronounced for bias voltages closer to short circuit, this leads to an underestimation of $J_{sc}/J_{gen| PL}$.

**Calibration based on reference solar cell**
The precise quantitative determination of the $J(V)$-image from PL requires especially the determination of the PL signal intensity of the camera detector if a PL of zero is expected which is for example influenced by leakage of light through the optical filter and detector noise, as discussed in the results section. Ideally, the setup should therefore be calibrated using a reference device for which $J_{sc}/J_{gen}$ is known.
In the following, the high performing (stabilized PCE = 21.2 %) perovskite cell (cf. Figure 2 and Figure S10) is employed to discuss the calibration of the setup. For the studied solar cell with a band-gap of 1.54 eV (805 nm) and an ideal EQE of 100 % between 300 nm to the band gap, the highest possible photocurrent under AM1.5g (ASTM G-173) is $J_{gen,max} = 27.6$ mA/cm². The determined stabilized $J_{sc}$ of $J_{sc}$ of 26.4 mA/cm² (cf. Figure S10) amounts to 95.7 % of this theoretical limit. Assuming an ideal photo absorption (maximum $J_{gen} = J_{gen,max}$), the highest $J_{sc}/J_{gen}$ is 0.96. Therefore, the sample could be used to calibrate the measurement setup. However, this approach is not fully satisfying for two reasons: first, it does not help to correctly assess the baseline of the PL measurements. Second, as shown in Figure S12, the device does not yield a completely homogeneous pattern for $J_{sc}/J_{gen| PL}$ which would mean that some areas need to be assigned with unphysical values larger than one.

**Calibration by baseline assessment**
Hence, a base-line correction for the PL setup was implemented by means of an alternative reference sample. Such a sample should induce the same parasitic signals in the detector but without the PL of the tested sample.
For the measurements of the III-V device, the baseline calibration was carried out by measuring the reflectance of the sample at a wavelength close to the microscope excitation wavelength and a measurement of the PL signal of the highly reflective front electrode metallization of the sample.
For the perovskite devices, a stack of glass / TCO / m-TiO$_2$ / ZrO$_2$ / carbon-graphite was fabricated to resemble a PSC without perovskite. This serves as a baseline to account for reflections and scattering from the different layers, whereas that the reflection from the (black) carbon-graphite layer can be assumed to approach zero. As shown in Figure S15, for very thin perovskite absorber layers, back-reflection of PL light from the gold electrode can have a significant impact on the measurement whereas for typical device thicknesses, the effect of optical reflection is over-compensated by a higher charge-recombination due to the presence of a metal electrode.
A second challenge lies in the choice of the most suitable light source. Ideally, the device should be illuminated by the AM1.5g solar spectrum. However, this is not possible as this spectrum overlaps with the PL spectrum of the solar cell. We investigated two LED light sources with a wavelength below the PL wavelength of red (632 nm) and blue (405 nm) light. Here, a trade-off needs to be made: The red light resembles the absorption depth of the AM1.5g spectrum acceptably well whereas the blue light is absorbed by the first few 100 nm of perovskite.[40] PL images recorded under blue light will therefore overestimate effects associated with processes of charge-carrier absorption and recombination at the front electrode. For the red light, however, the parasitic effect of the excitation light reflected from the sample and entering the detector is higher as there is a non-negligible overlap between the transmittance of the filter on the detector and the spectrum of the red LED.

Figure S10 shows the comparison of the PL($V_{oc}$) (a) and the $J_{sc}/J_{gen}|_{PL}$ image (b, c) under 632 nm and 405 nm LED illumination equivalent to the photon-flux of 1 sun, respectively. Notably, even after this baseline-correction, the $J_{sc}/J_{gen}|_{PL}$ averaged over the entire image for the two illumination sources was not identical. Under blue illumination it was estimated to 0.84 whereas for red illumination it was only 0.65.

Revealingly, the spots with the highest PL($V_{oc}$) intensity typically are reflected in the $J_{sc}/J_{gen}|_{PL}$ image by a region with a high $J_{sc}/J_{gen}|_{PL}$ that has a very low value in the center (cf. feature i). However, there are also features where this low center spot is not present, such as in ii).

In conclusion, the above discussion underlines the importance of accurate calibration and stresses that the quantitative analysis of the $J_{sc}/J_{gen}|_{PL}$ image needs to be carried out with care. Still, for a fast qualitative assessment of spatially distributed inhomogeneities, even a non-optimal setup is well suited, and the base-line correction can serve to assess the lowest limit of the optically determined $J_{sc}$. In the following, where we focus on qualitative comparison, the red LED is employed, as it resembles better the absorption under solar illumination.

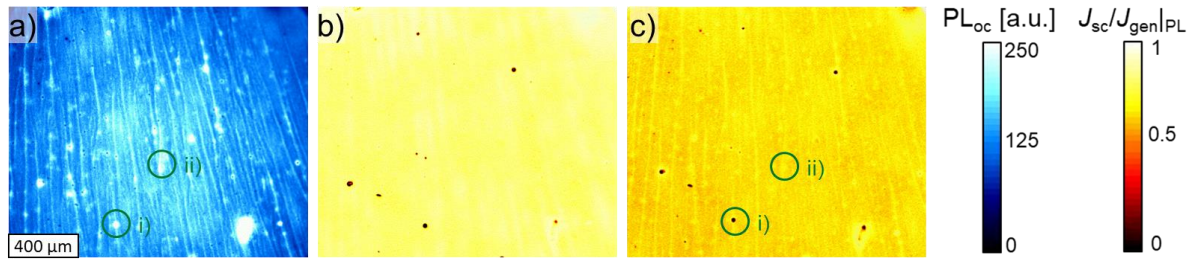

**Figure S12. a) PL($V_{oc}$) and (b, c) $J_{sc}/J_{gen}|_{PL}$ here determined from two PL images at open and short circuit of a high performing PSC. The device was illuminated with a blue (405 nm, b) and red LED (632 nm, c). Figure (a) and (c) are identical with Figure 2a and b and have been reprinted for comparison. The scale bar corresponds to 400 μm.**

## I. Further analysis of *PL(V)* and *I(V)* curves of the perovskite device

Figure S13 shows scaled representations of the local $J(V)/J_{gen}|_{PL}$ from Figure 3, normalized by the PL at short circuit. This representation can be used to compare the shape of the $J(V)/J_{gen}|_{PL}$ curves if the effect of different short circuit currents is ignored. Comparing Figure S13a with Figure 3c, it can be clearly seen that although the PL intensity of various spots reacts differently to the change of the bias voltage, $J(V)/J_{gen}|_{PL}$ follows the same exponential shape. This is can be further asserted by a semi-logarithmic representation, as shown in Figure S13b, displaying perfectly parallel and straight lines. The reason why the line is not continued for $V < 0.6$ V is probably due to a limited sensitivity of the PL camera at low changes of the PL intensity.

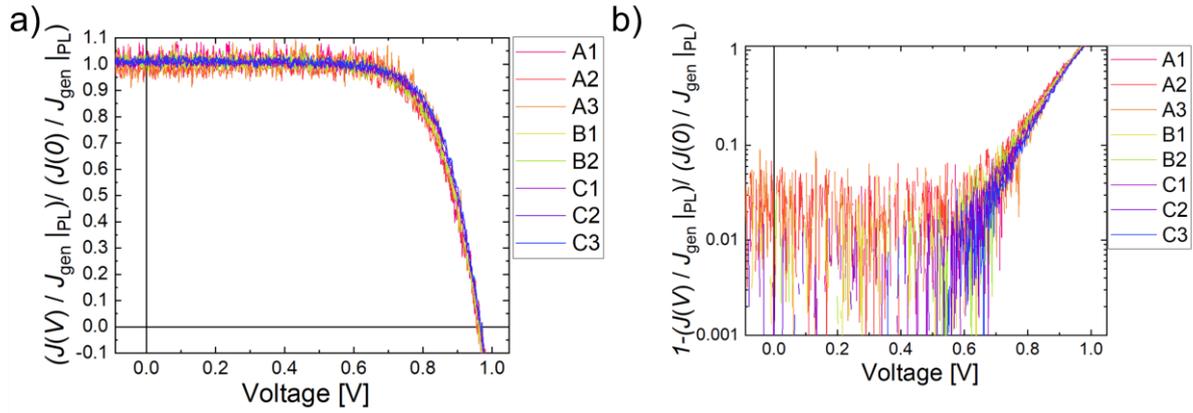

**Figure S13. a)** $J(V)/J_\text{gen}|_\text{PL}$ **curves of the spots discussed in Figure 3, scaled by the respective local short circuit current** $J(0)/J_\text{gen}|_\text{PL}$**. b) Semi-logarithmic representation, whereby the values from a) were subtracted from one.**

Figure S14a shows the average PL intensity (black curve) of the entire cell area measured during the voltage sweep displayed in Figure 3. For comparison, the current of the entire cell as measured in parallel is plotted (red). While the curves match up comparably well, it is interesting to note that the slope of the PL(V) curve is lower than that of the I(V) curve. From the considerations outlined below, one would rather expect the opposite if the cell was affected by a series resistance or by a non-ideal the diode ideality factor ($n_\text{id} > 1$). It appears that either the effects are compensated for by an unknown third effect or the two effects do not to affect the studied device.

In Figure S14b, the curves are plotted in a semi-logarithmic plot to allow a better comparison of the exponential shape of the curves. For this representation, the curves were scaled as follows. The current was scaled to $I_\text{scaled}(V) = (1 - I(V)/I_\text{sc})$ and the PL was scaled to $\text{PL}_\text{scaled}(V) = (1-(\text{PL}(V_\text{oc})-\text{PL}(V))/\text{PL}(0))$.

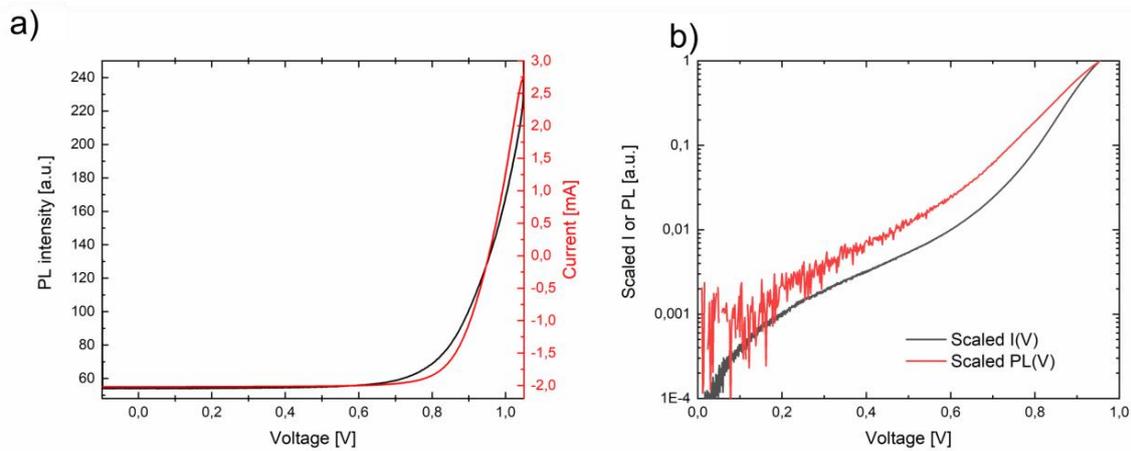

**Figure S14. a) Average PL intensity of the entire observed cell area (left axis, black curve) and the photocurrent (right axis, red curve) that was measured during the voltage sweep displayed in Figure 3. b) Scaled, semi-logarithmic representation of these curves.**

## J. Influence of the gold electrode on the PL($V_{oc}$) in perovskite devices

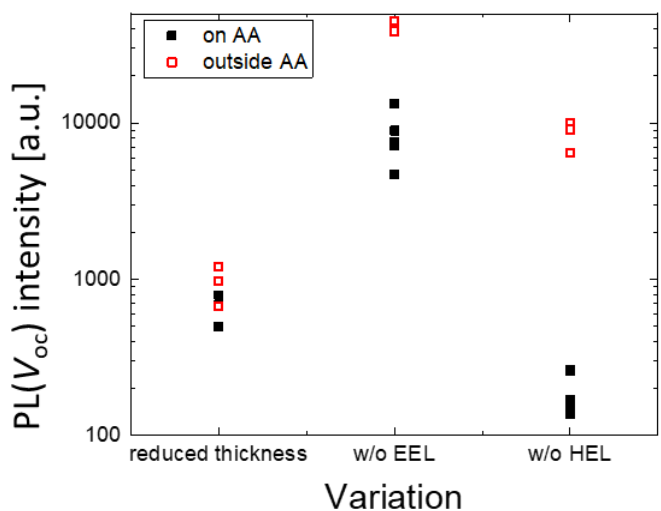

**Figure S15.** PL($V_{oc}$) intensity for a PSC range of samples with reduced thickness, without EEL and without HEL. The plot compares the PL($V_{oc}$) intensity measured on and outside the active area (AA), i.e., the region where the gold back electrode is deposited.

## K. Stabilized *I-V* parameters of further perovskite devices

Figure S16a shows the stabilized $V_{oc}$, as measured under a solar simulator, with the stabilized PL($V_{oc}$) measured under the PL microscope, for a range of samples with reduced thickness, without HEL or without EEL. The highest PL($V_{oc}$) is achieved with EEL-free devices, which show however only moderate stabilized $V_{oc}$. Hence, the trend of the PL($V_{oc}$) does not represent to observed $V_{oc}$. In contrast, the $J_{sc}/J_{gen}|_{PL}$ representation reflects much better the stabilized $V_{oc}$ as shown Figure S16b. This demonstrates once more that the PL($V_{oc}$) alone it is suited to estimate the device performance.

In Figure S16c, the stabilized $J_{sc}$ estimated from the solar simulator are compared with the $J_{sc}/J_{gen}|_{PL}$ estimated from the PL microscope. The dashed line represents the ideal case where $J_{gen} = J_{gen,max} = 27.6$ mA/cm². The measured values are overall in agreement with the expected, linear relation. Especially the values of the samples with reduced thickness lie below close to the dashed line (except for one outlier). The values of the devices without EEL and HEL lie above the dashed line. This could mean that $J_{gen}$ is below $J_{gen,max}$ due to non-ideal absorption or photogeneration. Especially for the HEL-free devices with very low photocurrents and PL intensities, the deviation could be due to the limitation of the method as discussed in the manuscript such as a non-ideal calibrated PL baseline.

Although we did not focus on the dependence of $V_{oc}$ and FF on the PL in this work, it is yet interesting to consider the show seemingly linear relationship between the stabilized PCE and $J_{sc}/J_{gen}|_{PL}$ as shown in Figure S16d.

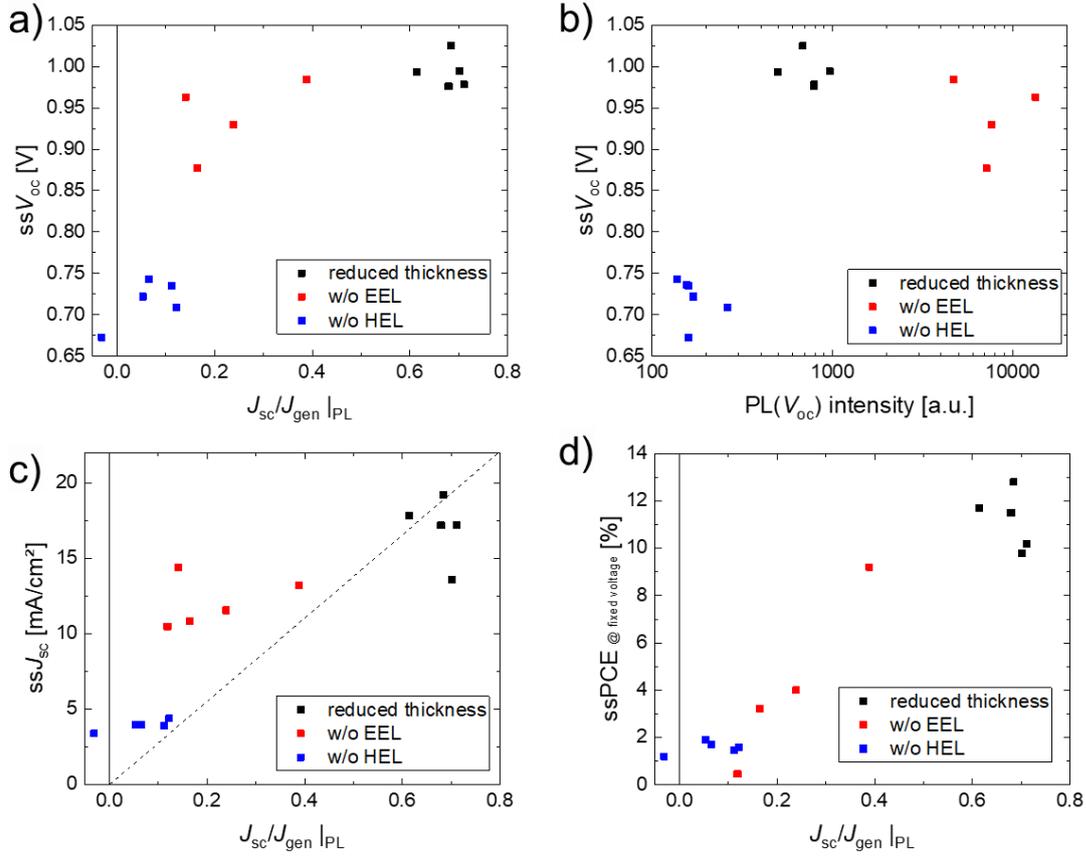

**Figure S16.** Stabilized $V_{oc}$, $J_{sc}$ and PCE (measured under at a fixed voltage close to mpp) as obtained from measurements at a class A solar simulator for a range of devices with a reduced perovskite thickness (red), without EEL (blue) or without HEL (magenta). The values are plotted against PL($V_{oc}$) or $J_{sc}/J_{gen}|_{PL}$, as obtained from a measurement at a PL microscope under red (623 nm) LED illumination.

## L. Spatial analysis of transient behaviour in perovskite devices

A further look at the local, transient effects of the spots in Figure 3 is presented in Figure S17. It shows the progression of the PL(V) intensity of these spots for different applied electrical bias, consecutively switching from $V_{oc}$ to maximum power point ($V_{mpp}$) again to $V_{oc}$, to short circuit (0 V), and again to $V_{oc}$, respectively with a dwell time of 60 seconds each. $V_{mpp}$ refers to the voltage at maximum power point as determined by a previous J-V sweep from $V_{oc}$ to 0 V.

Regarding the spots in groups A, B, and C, despite the difference in PL intensity, the transient reaction to a switch of the voltage bias is very similar. Again, the last switching is the most significant with a high overshoot of the PL intensity upon switching to open circuit, followed by rapid decrease within few seconds and a slower convergence to a steady-state value. It is instructive to compare the three different steps of $V_{oc}$: one initial step (0 < t < 60 s) before which the device has already been illuminated at open circuit for longer than five minutes. A second step (120 < t < 180 s) before which the cell was biased at $V_{mpp}$. Moreover, a third step (t > 240 s) that was preceded by a working point at short circuit. The effect of switching from $V_{mpp}$ to $V_{oc}$ is marginal and the initial steady-state value from the first step is reached to the greatest extent within few seconds. In contrast, there is a strong overshoot of PL when switching from short circuit to $V_{oc}$ in the third step. Here, the PL intensity does only stabilize and recover to the initial value for spots of group B while the PL is still decreasing after 60 s for spots of group C. The PL of spots of the group A increases during the process of switching – reaching PL intensities of group B in the final step - although it appeared to be stable in the first step at $V_{oc}$. Note that all spots exhibit a stabilized low PL upon switching to short circuit (180 < t < 240 s).

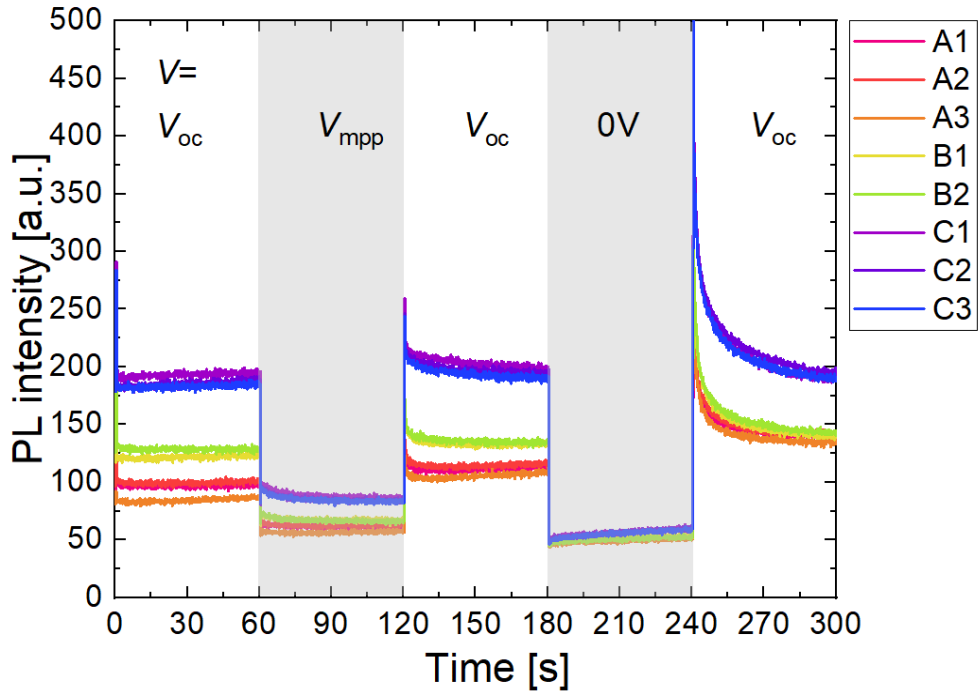

**Figure S17.** Transient evolution of the PL intensity at each spot of the sample depicted in Figure 3 for a successive bias of $V_{oc}$, $V_{mpp}$, $V_{oc}$, 0 V, and $V_{oc}$ for 60 seconds each.